# Electrophonic sounds from meteors and auroral audibility

Andrei Ol'khovatov

Independent researcher,

Russia, Moscow

https://orcid.org/0000-0002-6043-9205

email: olkhov@mail.ru

**Abstract:** Electrophonic sounds from meteors are sounds simultaneous with meteors (i.e. without delay). They are known for hundreds years. At least several hypotheses were put forward, but none of them received confirmation. Most hypotheses consider the energy of the meteoroid as a source of sound energy. In the early 1990s the author published several articles with another idea - a meteoroid is just a trigger of some processes which cause (in favourable geophysical conditions) electrophonic sounds. In this paper new arguments in favor of this idea are presented. Much attention is paid to sounds of aurora, which (in the author's opinion) could help better understand electrophonic sounds from meteors.

## 1. Introduction

Electrophonic sounds (from meteors) are sounds which appear prior to the acoustical disturbances generated by meteoroid motion in the atmosphere reaches the observational position. The term "electrophonic" appeared because it was a priori assumed by some researchers that the sounds were caused by electromagnetic radiation. A retrospective review can be found in (Keay, 1993), and in (Sung, et al., 2020) for the recent years.

Interest in studying electrophonic sounds increased in the 1980s - 1990s after publication of a series of articles by Professor Colin Keay. Keay proposed that these sounds are generated by a powerful source of ELV/VLF electromagnetic radiation from the trail of the bolide. The radiation propagates to the vicinity of the observer, and somehow transduces into acoustic waves (Keay, 1992, 1993, 1995, 1998). Here is what Keay wrote (Keay, 1992):

*"A VERY LONG-STANDING MYSTERY IN THE HELD OF METEOR fireball studies has now been satisfactorily resolved.*

*<...>*

*However, a viable physical explanation finally emerged in 1980 (Keay, 1980a), when it was shown that the plasma trail of a large fireball, or bolide, could generate Extra Low and Very Low Frequency (ELF/VLF) radio emissions (1 to 10 kHz). Then, whenever a suitable object happens to be close to an observer, sounds may be heard due to direct transduction of the electromagnetic energy into acoustic form. These are the anomalous fireball sounds, which with prescient accuracy, were sometimes called electrophonic sounds.*

*<...>*

*Figure 3 also clearly indicates a minimum bolide luminosity for sustained turbulent flow to be possible. The value is close to an absolute meteor magnitude of -9, in excellent agreement with the threshold luminosity for the generation of sustained electrophonic effects as determined empirically by Astapovich (1958) through his extensive investigations of such observations."*

Regarding the alleged mechanism of ELF/VLF radio emissions Keay wrote about the generation of electromagnetic waves in the meteoroid trail (Keay, 1998):

*"Calculations indicated that the abundant mechanical energy in the trail turbulence could be transferred to the geomagnetic field if the latter became trapped in the trail plasma. <...> When the trail plasma cools and recombines the released field energy excites ELF/VLF oscillations in the Earth-ionosphere cavity, as required to produce the electrophonic sounds."*

Concerning the process of the transduction Keay wrote (Keay, 1993):

*"Human electrophonic hearing (the direct perception of electrostatic fields varying at audio frequencies) has been reported (Sommer and von Gierke, 1964) but the field strengths required are large: several kilovolts per meter. Tests undertaken in an anechoic chamber with 44 volunteers to check their response at frequencies of 1, 2, 4 and 8 kHz showed quite wide variability between subjects (Keay, 1980c). The findings of Sommer and von Gierke were confirmed for the least sensitive subjects. At the higher frequencies, 4 and 8 kHz, the greatest sensitivity was shown by three*

*subjects (2 female) whose common characteristic was very loose or "frizzy" head hair. Their threshold peak-to-peak electric field strength was 160 V/m. Another subject (male) was found to be 3 to 4 db more sensitive at 2 and 4 kHz when wearing glasses.*

*<...>*

*Later, in another anechoic chamber, tests were conducted to test a number of mundane objects, including vegetation, for their ability to act as transducers (Keay and Ostwald, 1991). Under electric fields of 400 kV m$^{-1}$ peak-to-peak varying at 0.5, 1, 2 and 4 kHz, samples including aluminium cooking foil and typing paper produced sound levels in the 40 to 60 dB (SPL ref. 20 micronewton m$^{-2}$) range, while sprigs of casuarina pine and coastal myrtle produced from 10 to 25 dB (SPL). These represent minimal responses because the samples were not shaped or mounted in any special way to enhance their transduction ability. Of course, larger or more extensive amounts of the sample materials could be expected to produce similar sound levels at lower levels of electrical excitation. Furthermore, objects having resonant frequencies of vibration in the audio range would exhibit an enhanced response and color the sounds emitted. "*

It can be seen from the Keay's data that the minimum peak-to-peak electric field strength 160 V/m is needed to produce electrophonic sounds in the proposed Keay's mechanism. In 1984 Wang D.Y. with coauthors published an article (Wang, et al., 1984), in which it was shown that, from the mechanism proposed by Keay, it follows that the power of VLF/ELF radiation from the meteoroid trail should be on the order of $10^{11}$ - $10^{12}$ W or more. The authors of (Wang, et al., 1984) came to the conclusion:

*"The intensity of VLF emission required for the production of anomalous sounds necessitates unreasonably high currents. Furthermore, no long distance transmission of such VLF pulses is observed. We conclude that VLF emission from either meteor fireballs or aurorae cannot explain observations of anomalous sounds from these sources."*

In 1985 Keay published info about sounds reported during the Space Shuttle re-entry (Keay, 1985):

*"In 1984, Jim Oberg of Dickinson, Texas, and Drew Porter at the Johnson Space Center received numerous reports of people hearing a "swishing" sound as the Space Shuttle mission STS-51A reentered the atmosphere over Texas on November 16th. The same problem arises here as with meteors."*

After reading the information about the sounds accompanying the Space Shuttle flight, I wrote an article (Ol'khovatov, 1993), which showed that the mechanism proposed by Keay is not able to explain the sounds during the Space Shuttle flight (see details below). In the calculations performed by the author, the results of the article (Wang, et al., 1984) were used.

In his 1998 article Keay added some new info regarding the Space Shuttle (Keay, 1998):

*"When Space Shuttles began landing in Florida, people in Oklahoma, and Texas reported hearing them pass by as they re-entered the atmosphere.*

*Attempts were made to record the Shuttle sounds but were frustrated by last minute changes in flight plan. Then the reports ceased. There must have been some change in the Shuttle configuration or flight parameters. It remains a mystery."*

This info resulted in discussion between Keay and me on pages of "Meteorite!" magazine.  Since this magazine ceased to exist and became a bibliographic rarity, then this discussion is given below. My comments were published in the November, 1998 issue of the "Meteorite!":

*"Sir, I would like to comment Colin Keay's theory on electrophonic sounds (Meteorite!, August '98). At first, Keay's idea means that a bolide producing electrophonic sounds is to generate in its wake VLF radiation with the power at least about $10^{12}$ W (Wang, D.Y., et al. J. Roy. Astron. Soc. Can. v.78, no.4, p.145 (1984)). As far as I know, no present theory either experiments with turbulent ionized wakes or predicts such superpowerful VLF radiation. Anyway, if it is realized somehow, it leads to spectacular effects, for example, to enormous Joule heating of the wake (due to extremely large electric currents) transforming the wake into object as bright (seen from the ground) as at least the Sun (see: Ol'khovatov, A.Yu. Geomagnetism and Aeronomy. English ed., v.33, no.2, p.264 (1993)). And, of course, this "super-radiation" would produce remarkable global effects - but none of them are known.*

*Moreover, the "electrophonic" bolides look like "ordinary" bolides - no signs of superpowerful electric currents in their wake (for example, see published in the Keay article the photo of the Peekskill fireball).*

*By the way, in many electrophonic events the power of aerobraking,*

*i.e. the power deposited by a meteoroid into the atmosphere, which is to be the energy source of the proposed VLF radiation was much less that the power of the latter.*

*For example, on p.108 of December 1993 Sky & Telescope there is a photo of a "hissing" fireball of April 15, 1993, which became visible at a height of about 98 km and disappeared at 75 km after a flash with magnitude -9.5. Its looked like an "ordinary" bolide - no hints on the "superpowerful radiation" from it (I don't even touch here that it was too high for the existence of a "turbulent wake").*

*But the best example is the Space Shuttle re-entry, accompanied with electrophonic sounds. At that time the Shuttle was about 60-70 km high with a speed about 5-6 km/s (seen from the ground as an object with magnitude -1 to -2). Its aerobraking power was about $10^{10}$ W. Moreover in its photo fortunately taken about the same time (see Aviation Week & Space Technology v.122, no. 3, p.85 (1985)) again there is no evidence of the "super-radiation".*

*I'd like also to add that, of course, some weak plasma instabilities in the wake under the action of ionospheric factors etc. can take place, anyway they are negligible compared with the "super-radiation".*

*Electrophonic phenomena have been observed in cases when it is known reliably that intense VLF radiation from the fireball wake did not occur. And I could add that if it takes place during the Shuttle re-entry, it would produce a devastating effect on its avionics and possibly even the crew and its fuselage!*

*By the way, Keay's remark that "the reports [on the Space Shuttle electrophonic sounds] ceased", also hints that the physical agent of the sounds was not produced by the Space Shuttle wakes, but more probably just*

*was triggered by the flights in certain geophysical circumstances (see for details: Ol'khovatov A.Yu. Izvestiya [Russ. Acad. Sci.], Physics of the Solid Earth. English edition, v.29, no. 12, p.1043 (1993) and Priroda, no. 5, p.68 (1995) (in Russian)).''*

This my letter resulted in reaction by C. Keay, and here is the follow-up discussion with him in the February 1999 issue of "Meteorite!" (p.42):

*''*

### *Electrophonic Sounds Debate*

*Sir, A. Ol'Khovatov (Letters, Meteorite!, November '98) asserts that bolides must generate "superpowerful radiation" for electrophonic sounds to be heard by witnesses on the ground. This is not a requirement of my theory for the production of electrophonic sounds from large bolides. Ol'Khovatov bases his claim of terawatt source power on figures published by Wang, et al.. (JRASC, 78, 4, 145-150, 1984) on the basis of electric field strengths of 160 V/m measured in my laboratory tests of direct human sensitivity without transduction by external artifacts (JRASC, 74, 5, 253-260, 1980). These fields happen to be in complete accord with earlier work by Ivanov and Medvedev (Geomag. and Aeronomy, 5, 216-219, 1965) who showed that comparable induced field variations could result from some bolides in appropriate trajectories when the Earth's geoelectric field is higher than usual. On the question of power levels, however, I consider as much more realistic Vitaly Bronsten's calculation (Solar Sys. Res., 17, 70-4, 1983) of megawatt source power based on my theory for the generation of the kilohertz radio emissions which give rise to observers hearing sustained*

*electrophonic sounds.*

*On the question of energy requirements, it may be worth noting that in the early days of radio broadcasting, stations with a radiating power of only a kilowatt at distances of hundreds of kilometers could be received by unpowered crystal sets! Given an appropriate transducer the human ear is able to detect extremely low levels of power under quiet conditions.*

*Returning to the problem of electrophonic sounds from bolides, it is becoming apparent that there are many mechanisms by which ELF/VLF energy may be transmitted instantly to an observer, depending on the situation. Electromagnetic energy may be produced by explosive disruptions of a bolide (similar to the transient EM pulses from a nuclear air-blast), or by my mechanism (Science, 210, 11-15, 1980). Also there are purely electrostatic mechanisms such as that of Ivanov and Medvedev or simply the action of a zenithal meteor entering on a steep trajectory to produce a conductive path below the ionosphere which creates a strong transient in the geoelectric field at the surface of the Earth. This latter effect could explain the recent aural observations of Leonid fireballs by K Matsuura in Japan (private communication).*

*Clearly, much further study is needed to clarify outstanding problems of geophysical electrophonics.*

*Colin Keay*
*Physics Dept.*
*University of Newcastle*
*NSW, Australia*

*Ol'Khovatov replies...*

In response to my criticism of C. Keay's theory about the origin of the electrophonic sounds, C. Keay states that just a megawatt power source of VLF/ELF radiation from a bolide's wake is enough to produce electrophonic sounds. But I can't accept his arguments.

The terawatt power source estimation is based on applying well-known formulas of electromagnetic radiation to Keay's idea. So to go down from "terawatts" to the "megawatts", C. Keay must find a mistake in Wang D. et al.'s calculations or change something in his theory. These are not done.

V. Bronsten was just trying to calculate the magnetic energy that could be stored in a bolide's wake according to the Keay theory and he got the energy relaxation rate estimation of "gigawatts". Then he speculated that about 0.001 of the energy could be transformed into radio waves. In other words, if his result is correct, it just imposes an upper limit on the power of the hypothesized radio waves - on the order of megawatts. As we just saw, it is a million times less than needed to produce the required 160 V/m electric field strength on the ground.

By the way, if the Keay "megawatt" estimation (which completely contradicts his 160 V/m value) is correct, that means that residents dozens of miles away from VLF/ELF radio stations (many of them have about such power) must hear the electrophonic sounds around the clock! And what about the radio station's personnel?

In my previous Letter, I also have mentioned several other evidences against the Keay idea. Even the shift from "terawatts" to "megawatts" is not able to avoid the majority of them. For example, even "megawatt" power can produce remarkable effects on Space Shuttle avionics. And, as I have

*written, the electrophonic sounds were heard when a rather weak bolide was too high for the existence of a turbulent wake. I hope that C. Keay will consider these aspects too, if he insists on his theory.*

*Andrei Ol'khovatov,*
*Moscow, Russia*

*In reply...*

*Ol'Khovatov still does not realize that detection of electrophonic sounds from fireballs relies on the presence close to the observer of a suitable transducer of adequate efficiency (hence my crystal set analogy). That is why observations of such sounds are rare and quite capricious. Time will tell who is correct. Right now an explanation for the phenomenon better than mine has yet to emerge. My explanation has thus far met most criticisms except those based on misinterpreted data."*

*Colin Keay*

*Editor's note: 1 megawatt = 1 million ($10^6$) watts, 1 gigawatt = 1 billion ($10^9$) watts, 1 terawatt = 1 trillion watts ($10^{12}$) watts. Further research involving the Space Shuttle to resolve this matter is being planned. Stay tuned."*

Despite that the discussion was stopped by the editor, in my opinion its key points were presented. The main key point against the Keay's mechanism *("abundant mechanical energy in the trail turbulence could be transferred to*

*the geomagnetic field if the latter became trapped in the trail plasma. <...>*
*When the trail plasma cools and recombines the released field energy*
*excites ELF/VLF oscillations...")* is that it can't generate electromagnetic
radiation powerful enough to create electric field at least 160 V/m near an
observer in real cases of reported electrophonic sounds.

Since that time more and more arguments appear against the Keay's
"geomagnetic-field-relaxation" mechanism. So several researchers made in
2017 a revival of a photoacoustic mechanism, writing in its favor (Spalding
et al., 2017):

*"Prior to now, the means by which energy from meteors could be propagated*
*to Earth and then converted into audible sound has not been adequately*
*explained and validated by experiment. <...> A previous hypothesis of*
*coupling to natural antennas from RF radiation generated by plasma*
*oscillations[1,2] does not seem to be adequately supported by observational*
*evidence of radio waves emanating from meteors[12–15]."*

Nevertheless, the appearance of the mechanism proposed by Colin
Stewart Lindsay Keay (1930 - 2015) played a large positive role in the study
of electrophonic phenomena. Keay's articles, his activity contributed greatly
to the fact that many scientists turned to electrophonic research, and many
important results were obtained.

An interesting mechanism was proposed in (Kovalyova, 2018)
(translated from Russian by A. Ol'khovatov):

*"At altitudes of about 100 km or less the movement of a meteoroid is accompanied by a glow, including in the ultraviolet range. This leads to additional ionization and excitation of ambient air. When large meteoroids fall, excitation by light of air molecules (electronic, vibrational, rotational) reaches to the surface of the Earth. When streams of small meteoroids (the so-called Leonids, Perseids, etc.) fall, the air excitation reaches only stratospheric heights. Assuming that the source of noise accompanying the fall of meteoroids is the interaction of an electromagnetic wave with light air-ions in an acoustically active medium (containing nonequilibrium vibrationally excited molecules), a model for describing such an interaction was built."*

However it predicts the delay between faint meteors and their electrophonic sounds - from (Kovalyova, 2018) (translated from Russian by A. Ol'khovatov):

*"Falls of large meteoroids are accompanied by significant modification the entire atmosphere, the accumulation of vibrational excitation at all heights, which, apparently contributes to the excitation of electrophonic noises both in stratosphere, and at the surface of the Earth at the frequencies of all light ions present in the atmosphere. In addition, the cyclotron frequencies with the fall of large meteoroids shift with the simultaneous excitation of a large-scale magnetosonic wave, leading to fluctuations in the geomagnetic field. Similar to electrophonic noise signals, with a delay of several tens of seconds, can be also expected from small meteoroids (meteor showers of the Leonids, Perseids and etc.)."*

In reality the delay is absent.

## 2. Arguments for a geophysical factor

The author of this paper inclines to think (the author promotes this idea since early 1990s) that a meteoroid triggers in favorable geophysical circumstances some processes which produce electrophonic sounds. In other words, most of the energy is supplied from geophysical environment. In (Ol'khovatov, 1994) some arguments were presented. Here are some newer arguments in this favor in chronological order.

**- Observations of a sprite in coincidence with a meteor-triggered jet event (Suszcynsky, et al., 1999).** The event consisted of three stages: (1) the observation of a meteor, (2) the development of a sprite in the immediate vicinity of the meteor as the meteor reached no lower than about 70 km altitude, and (3) a slower-forming jet of luminosity that appeared during the late stages of the sprite and propagated back up the ionization trail of the meteor. In the opinion of the authors (Suszcynsky, et al., 1999):

*"In summary, at this point in the analysis the occurrence of the sprite and associated positive CG discharge in spatial and temporal proximity to the meteor cannot be conclusively shown to exhibit a causal relationship. The occurrence of the ensuing jet event is clearly related to the meteor's ionization trail, and although the initiating mechanism is not obvious, some similarities between these observations and previous me-*

*teor observations are apparent."*

Interestingly that the meteor was rather faint - its brightness was conservatively estimated at third magnitude.

**- Discovery that faint meteors can be accompanied with electrophonic sounds.** Here is a result of statistical analysis of the reported electrophonic sounds for (relatively) faint meteors (brightness magnitude from -1 to -5) and bolides (brighter than -10) from (Vinković, et. al., 2002):

Faint meteors: crackling, sizzling like bacon frying, sizzling "sss", soft hissing, hissing followed by a crack, ffffp, short burst of static, short sharp crack, broken filament shaking in blub, "thwuck", pop, crackling, swoosh, woosh, high pitched whistle, fizzing with crackling, loud high-pitched hissing, faint hissing, crackled/hissed, pop.

Bolides: whistling with buzzing, whisper, sizzling, rustling like a rocket, wood on fire, white noise, thrumming, lit match, "sss" followed by pop, "voom", pop, whoosh like rustling, hissing/fizzing.

It can be seen that despite difference in brightness in order of $10^4$ times (and associated difference in energy deposition) there is not much difference the electrophonic sounds.

By the way the photoacoustic interpretation (Spalding et al., 2017) can't explain electrophonic sounds from meteors fainter than about -12 magnitude as according to the interpretation (Spalding, et. al., 2017): *"This*

*suggests that an observant person in a quiet environment containing good transduction materials could hear photo-acoustically induced sound from a − 12 magnitude or brighter fireball — assuming it emits light modulated at acoustic frequencies."*

**- The first recording the electrophonic sounds.** The authors of (Zgrablic, et al., 2002) presented the first instrumental detection of electrophonic sounds obtained during the observation of 1998 Leonids from Mongolia. Two Leonid fireballs of brightness -6.5$^m$ and -12$^m$ produced short, low-frequency sounds, which were simultaneously recorded by microphones in a special setup and heard by different observers. Simultaneous measurements of electromagnetic ELF/VLF radiation above 500 Hz did not reveal any signal correlated to the electrophonic event.

**- Sounds of aurora were recorded.** Stories of the aurora sounds (i.e. sounds which accompany movements of aurora without delay) are known for centuries (Silverman and Tuan, 1973). But the first time they were recorded in Finland in 2011. Here is from (Laine, 2012):

*"Twenty-one short sound events, clearly discernible from the background noise, were detected afterwards in the recording. The best ten were selected for closer analysis. These were the loudest events (claps) and exhibited relatively high mutual cross-correlations and a low background noise. Three of these events were detected with all three microphones. This is the first time during our studies when similar sound events were recorded simultaneously with multiple microphones during a geomagnetic storm. Knowing the separation distance of the microphones (B&K vs. Zoom) it was*

*possible to estimate the direction of the sound source. The synchronously recorded VLF signal provided data to estimate the distances of the sound sources assuming that a certain magnetic field event is associated with a certain sound event. Three independent methods to estimate distance to the sound sources yielded an average value of 70 meters. This outcome, together with the direction estimate, indicates that the average altitude of the sound sources measured from ground was only 60-70 meters."*

Later a mechanism was proposed - from (Laine, 2018):

*"The sound sources were localized about 75-80 meters above the ground in the temperature inversion layer. The layer is formed during calm evenings and nights and it collects negative ions from the ground and positive from the upper atmosphere. These two layers with opposite charges are relatively close to each other, but due to the lack of any vertical movements in the air around the layers, they are not in contact but stay charged and stabile [4]. During a geomagnetic storm with bright and lively aurora a rapid change in the geomagnetic field may trigger a discharging process causing both MFPs and after a delay audible clapping or crackling noise on the ground."*

Additional analysis hints on possible relation of the auroral sounds with Schumann resonances (Laine, 2019):

*"The focus of the present paper is in the acoustic analysis of the crackling sounds that are sequences with complex temporal structures. It is shown that the sounds share similar rhythmical patterns, which are*

*connected to the electromagnetic resonances in the atmosphere, also known as the Schumann resonances."*

Later auroral sounds were recorded also by Japanese researchers (Amo, et al., 2020).

**- The 2013 Chelyabinsk bolide.** The 2013 Chelyabinsk bolide was brighter than Sun, and was accompanied with electrophonic sounds (Popova, et al., 2013). According to (Popova, et al., 2013) from the 1,674 people interviewed during the internet survey, 198 reported hearing sounds. The sound effects were described as hissing, as if you run fireworks noise interference, the sound of bengal light, crackle, sparking, crackling, rustle, rustling, like a whistle, squeaking, rumble, and the sound of a passing plane. Here are some examples from (Popova, et al., 2013):

*"While in his office in Yemanzhelinsk, Evgeny Svetlov, the head of the Yemanzhelinsk administration and an electrical engineer by training, heard a noise like the buzz of the electrical transformer during the main bolide flash. Alexander Polonsky, a car driver, heard a noise like the roar of two fighter planes even before he saw the bolide, while standing on a street in Yemanzhelinsk. Finally, Vladimir Bychkov, a police programmer and physicist by training, heard a noise like the sizzle of oil in a frying pan, during the bright stage of the bolide while he stood on a square in Chelyabinsk. The noise appeared to be from the direction of the bolide. The noise stopped at the main bolide flash, but there was a short sound like a clap during the flash. None of these witnesses were wearing glasses."*

Remarkably that according to (Popova, et al., 2013) two local electricity companies in Yemanzhelinsk district reported that there were no significant voltage surges across power lines at the time of the bolide. The electricity supply did not switch off.

The 2013 Chelyabinsk bolide was (roughly) 10 billion times brighter than a faint meteor with electrophonic sounds, but there is not much difference in manifestation of the sounds in their character, and spatial coverage. It looks like electrophonic phenomena demonstrate a kind of "saturation".

## 3. Discussion

The abovegiven facts indicate that the energy release of the meteoroid plays, rather, the role of a trigger in the generation of electrophonic sounds. The difference between a faint meteor and a bright bolide is mainly in possibility (probability) to trigger the sounds. Brighter the meteor - larger the probability.

It is quite probable that the process of generation of electrophonic sounds of meteors is similar to the mechanism of generation of recorded sounds of aurora. In (Laine, 2019) it is assumed that the generation mechanism is associated with corona discharges in a temperature inversion layer ~ 70 m above the ground. Also it was shown that the sounds share similar rhythmical patterns, which are connected (in the Laine's opinion) to the electromagnetic resonances in the atmosphere, also known as the Schumann resonances. If confirmed then it hints that global processes are involved.

Let's consider this issue from a slightly different angle. It is known that the precipitation of energetic particles into the ionosphere can lead to a significant (several times or more) change in the near-the ground electric field - see, for example, (Shumilov, et al., 2015). A discussion of the reasons for this phenomenon is beyond the scope of this article, in this case the very fact of this phenomenon is important. In (Silverman and Tuan, 1973), data from measurements of the electric field are given, which showed the possibility of its significant enhancement during the aurora:

*"The electric field measurements of Olson were taken with field mills at Duluth and Minneapolis. The instruments were suspended some twenty meters above the ground level. The measurements at Duluth showed very sharp peaks with peak values close to 10,000 V/m and rise times of about 2 to 4 minutes. <...> He reports that the effects are observed only when an intensely active coronal form, in subauroral latitudes, and the westward surge form, in the auroral zone, are in the vicinity of the magnetic zenith. At subauroral latitudes he notes that $K_p$ is generally greater than 5 when effects are observed, and that the weather is generally clear and calm. No effect is noted in clear weather without aurora. Finally, he finds that under favorable conditions an event is observed in about one out of four cases. <...> Mühleisen (1969) has noted that at the IUGG Conference in 1967, Olson's results and objections thereto were discussed, and no errors in the measurement were found.*

*The conditions noted as favorable for high electric fields here are comparable to those which, as indicated, favor the occurrence of auroral*

*sound events. Taken all together, and making allowance for the correlations shown in Section 3, these results are consistent with the hypothesis that auroral sound events are produced by brush discharges produced by transient high potential gradients associated with intense overhead aurora."*

Thus, in some cases, at an altitude of a couple of tens of meters, the magnitude of the electric field during the aurora can reach rather significant values. In such conditions peculiar phenomena could take place as, for example, reported by a Russian geologist Popov L.N., and which was presented in translation into English form in (Ol'khovatov, 2020). The geologist was "attacked" by buzzing and whistling invisible flying "arrows".

As for the specific mechanism of sound formation, possibly it would be reasonable to consider the one similar to discussed in (Kovalyova, 2018). Under conditions of a sufficiently strong electric field, the vibrational degrees of freedom of molecules and ions can be excited. In a number of works, for example, in (Perelomova, 2017; Zavershinskii, et al., 2013; Molevich, et al., 2013), the possibility of excitation of an acoustic pulse due to the transfer of energy from the vibrationally excited state of molecules to an acoustic signal was shown.

The author is not aware of the data on measurements of the near-surface electric field in association with with electrophonic meteors (or bolides). However, in a place located almost 1,500 km west from the 2013 Chelyabinsk meteoroid fall, the increase of the near-surface electric field started in a minute after the Chelyabinsk meteoroid entered the atmosphere (Spivak, et al., 2019).

One way or another, the question remains - how can an aurora, and even a faint meteor, lead to the appearance of a strong electric field near the earth's surface? When discussing possible answers, it makes sense to consider the following. First, we are dealing, after all, with rather rare phenomena, especially in the case of faint electrophonic meteors. Secondly, the answer is likely to be found in the direction of investigating functioning of the global electrical circuit. But of course the discussed mechanism of electrophonic sounds generation which is associated with increase of the near-surface electric field is still speculative. The existing data is too sparse for any solid conclusion. Probably it is still too early to pinpoint reliably the physical mechanism.

The author proposed (Ol'khovatov, 1994) a role of endogenic factor in generation of electrophonic phenomena. It is known that the spatial distribution of auroras reacts to the features of the earth's surface, in particular, to the coastline and geological structures (Popov, et al., 1989, 2010). The specific physical mechanism of this influence has not been unequivocally established. In addition, one can cite the case considered in (Ol'khovatov, 2020), which happened on May 25, 1989, near a village 109 kilometers from Budapest (Hungary). A man was killed by something which resembled an "electric discharge from underground" in none-thunderstorm situation.

It is noteworthy that in (Kochemasov, 2003; Zherebchenko, 2004) it was reported about the identification of the relationship between the spatial distribution of the frequency of reports about electrophone bolides with large-scale geological structures. But these works are just the first steps. It is advisable to continue such studies with larger statistical data.

It is reasonable to mention that some geophysical phenomena could be associated with a flying luminous objects and sounds resembling the electrophonic from meteors - see some examples in (Ol'khovatov, 2020). These phenomena should not be confused with meteors with electrophonic sounds.

## 4. Conclusion

1) There is not much difference between electrophonic sounds from a faint meteor and a bright bolide. The difference is mainly in probability of appearance of the sounds.

2) The latter factor is one more argument in favor of the author's idea that a meteoroid is not a main source of energy for generation of electrophonic sounds, but sooner is a trigger/regulator of some processes in geophysical environment. These underlying physical processes are to be identified still.

3) In the author's opinion much more attention should be paid to investigating role of engogenic processes in the global electric circuit.

4) As field research of the meteor's electrophonic sounds is associated with "happy chance" factor, then researching sounds of aurora could be fruitful for better understanding electrophonic sounds from meteors.

The electrophonic sounds are unexplained still, so science should continue to investigate them...

**ACKNOWLEDGEMENTS**

The author wants to thank the many people who helped him to work on this paper, and special gratitude to his mother - Ol'khovatova Olga Leonidovna (unfortunately she didn't live long enough to see this paper published...), without her moral and other diverse support this paper would hardly have been written.

Supplementary Materials.